\documentclass[a4paper]{article}
\usepackage[caption=false,font=footnotesize]{subfig}
\usepackage{INTERSPEECH2022}
\usepackage{hyperref}
\usepackage{enumitem}
\usepackage{multirow}
\usepackage{multicol}
\usepackage{numprint}
\usepackage{comment}
\usepackage{amsmath,amsfonts}
\usepackage{amssymb}
\usepackage[font={small,it}]{caption}
\title{Analyzing Language-Independent Speaker Anonymization  \\ 
Framework under Unseen Conditions}
\name{Xiaoxiao Miao$^1$, Xin Wang$^1$, Erica Cooper$^1$, Junichi Yamagishi$^{1}$, Natalia Tomashenko$^2$}
\address{
$^1$National Institute of Informatics, Japan 
 $^2$LIA, University of Avignon, France}

\email{xiaoxiaomiao@nii.ac.jp}

\begin{document}

\maketitle

\begin{abstract}
In our previous work, we proposed a language-independent speaker anonymization system based on self-supervised learning models. Although the system can anonymize speech data of any language, the anonymization was imperfect, and the speech content of the anonymized speech was distorted. This limitation is more severe when the input speech is from a domain unseen in the training data.
This study analyzed the bottleneck of the anonymization system under unseen conditions. It was found that the domain (e.g., language and channel) mismatch between the training and test data affected the neural waveform vocoder and anonymized speaker vectors, which limited the performance of the whole system. Increasing the training data diversity for the vocoder was found to be helpful to reduce its implicit language and channel dependency. Furthermore, a simple correlation-alignment-based domain adaption strategy was found to be significantly effective to alleviate the mismatch on the anonymized speaker vectors. Audio samples\footnote{\url{https://xiaoxiaomiao39.github.io/IS2022-SAS/}} and source code\footnote{\url{https://github.com/xiaoxiaomiao39/SSL-SAS}} are available online. 
\end{abstract}

\begin{figure*}[hbt]
\centering
\vspace{-1.4cm}
\subfloat[][Training procedure]{\includegraphics[trim=0 12 0 0, clip,height=5.8cm]{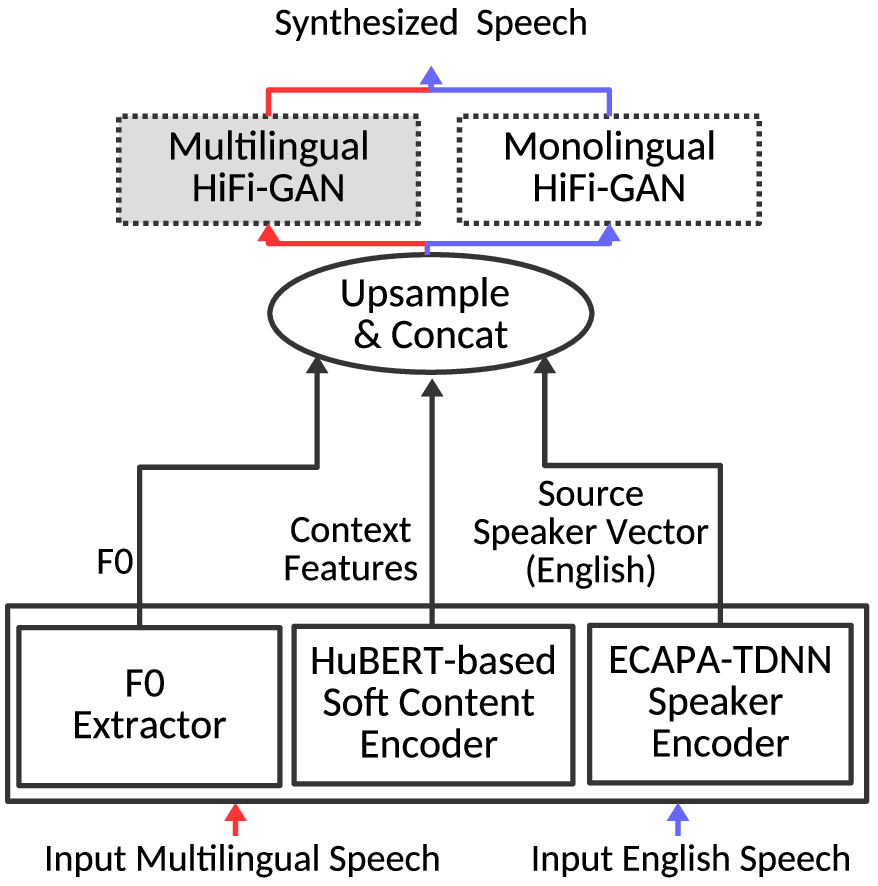}} \hspace{1.5cm}
\subfloat[][Anonymization procedure]{\includegraphics[trim=0 12 0 0, clip,height=5.8cm]{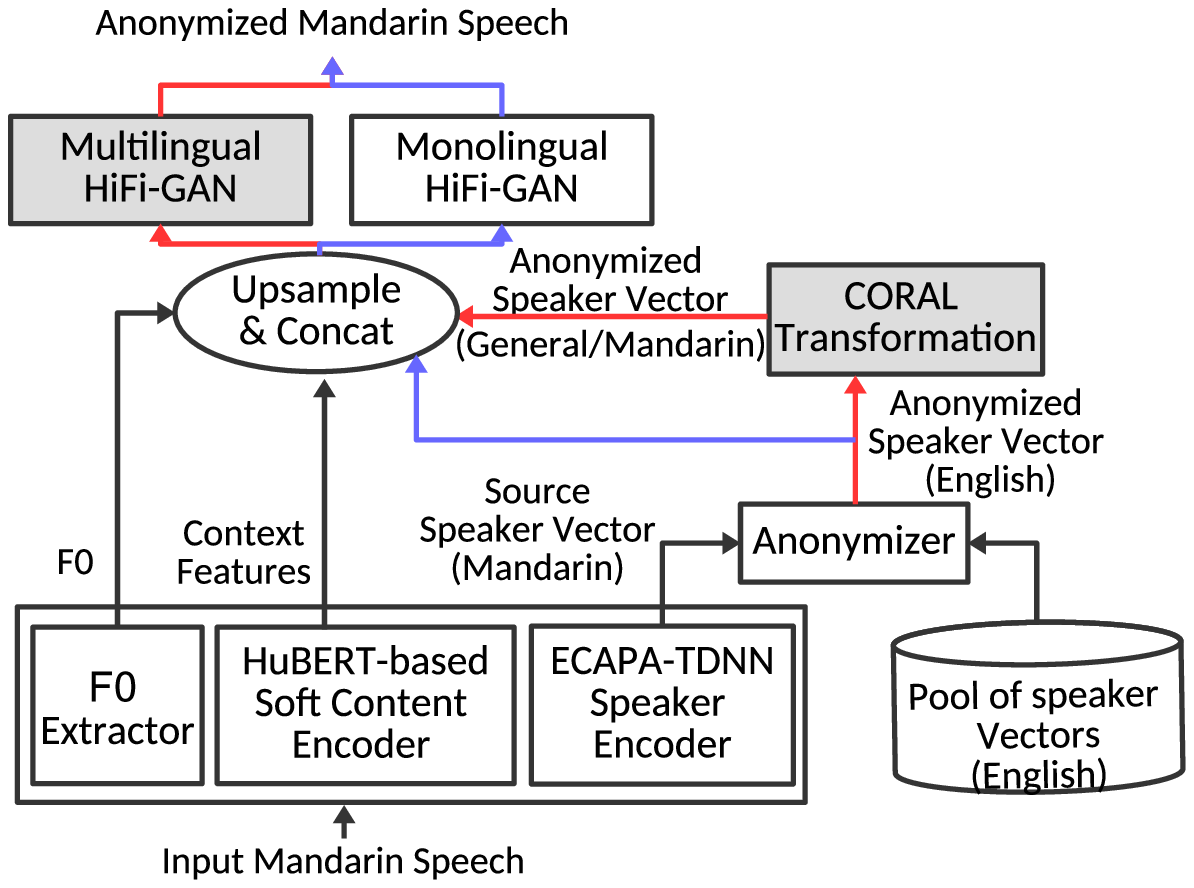}}\\
 \vspace{-0.3cm}
\caption{Diagram of language-independent SAS. Blue and red arrows indicate components investigated in this study.}
\label{fig:structures}
 \vspace{-0.6cm}
\end{figure*}

\section{Introduction}
The human voice contains a wealth of personal information, such as the speaker's identity and emotion. Since personal information can be revealed by advanced speaker or other types of recognition systems, the demand for privacy-preserving technologies is growing.
Although there is no legal definition of privacy \cite{Nautsch2019}, through the initiative called VoicePrivacy Challenge (VPC) 2020
\cite{tomashenko2020introducing, tomashenko2021voiceprivacy}, the research community has defined a speaker anonymization task, the goal of which is to protect the speaker identity information (privacy) while maintaining the speech intelligibility and naturalness (utility). 

While other related methods exist \cite{jin2009speaker}, a recently proposed deep neural network (DNN)-based  \cite{fang2019speaker} method was used as the VPC 2020 primary baseline to disentangle speaker and other information in the speech data and synthesize speech after anonymizing the speaker information. The effectiveness of this speaker anonymization system (SAS) has been confirmed on English test sets. However, it requires large amounts of text transcriptions for English training data to obtain accurate linguistic representations, 
which makes it impossible to use for an unknown language.
While other digital-signal-processing-based methods \cite{patino2020speaker} require little or no training data, they are less effective than the DNN-based methods at protecting the speaker identity \cite{srivastava2020evaluating, tomashenko2021voiceprivacy}.

Aiming at an effective speaker anonymization solution that can be applied to the speech of any language, we have proposed a self-supervised learning (SSL)-based language-independent SAS  \cite{miao2022language}. %
It uses an SSL-based content encoder to extract general context representations regardless of the language of the input speech. The whole SAS requires no text labels or other language-specific resources, allowing the system to anonymize speech data from any language. 
This SAS has been applied to Mandarin speech data. Even though the Mandarin language was completely unseen to the SAS, the Mandarin speech samples were anonymized reasonably well. However, it was also observed from an increased character error rate (CER) that the speech contents were distorted after anonymization. While the trade-off  between speaker anonymization and speech intelligibility is common to many SASs \cite{tomashenko2021voiceprivacy}, our goal is to push the limit of the language-independent SAS and improve both privacy and utility metrics in unseen conditions. 

This study takes one step towards our goal by experimentally analyzing the performance bottleneck of the SAS. Specifically, it analyzes how the components such as the speech generator (i.e., vocoder) are implicitly dependent on a particular language or channel in the training database of that language. While keeping the target language (i.e., Mandarin) unseen, this study finds it beneficial to increase the language diversity of the training data, for example by adding German, Italian and Spanish speech data. 
Furthermore, this study investigates the language/channel mismatch brought by the impure speaker identity representation. 
It is found that the mismatch can be alleviated by transforming the anonymized speaker vector using a simple CORrelation ALignment (CORAL)-based domain adaption strategy \cite{sun2016return}.
Transforming the anonymized speaker vector from the source domain (English) to a general domain covering German, Italian and Spanish data achieves better performance. When transforming to a more ideally matched domain with a few samples of Mandarin data, the CORAL provides the best utility with remaining high privacy.
These findings are expected to be useful to the community for building a better language-independent SAS that can work under unseen conditions.

\section{SSL-based Language-Independent Speaker Anonymization System}
\label{sec:ssl-sas}
The baseline SSL-based language-independent SAS \cite{miao2022language} disentangles speech into the fundamental frequency (F0), speaker identity representation, and content representation. Figure \ref{fig:structures} shows the training and anonymization procedure of the baseline system (black and blue arrows path), respectively.
There are two steps in the training stage:

\noindent
\textit{1) Original F0, speaker identity, and context feature extraction from the original speech recordings.}
The YAAPT algorithm \cite{kasi2002yet} is used to extract F0.
The emphasized channel attention, propagation and aggregation in a time delay neural network (ECAPA-TDNN) speaker encoder trained on the \textit{VoxCeleb-1 \& 2} \cite{nagrani2017voxceleb, chung2018voxceleb2} datasets is used to extract 192-dimensional speaker identity vectors.
To extract finer-grained context representations, the HuBERT-based soft content encoder \cite{van2021comparison} downsamples the input speech into a sequence of 768-dimensional continuous representations, which is reduced to 200 dimensions through linear projection. Note that this HuBERT-based soft content encoder took the CNN encoder and the sixth transformer layer from the input of a HuBERT Base model \cite{hsu2021hubert} pre-trained on LibriSpeech\footnote{\url{https://github.com/pytorch/fairseq/tree/main/examples/hubert}} as the backbone. It was fine-tuned on the  \textit{LibriTTS-train-clean-100} \cite{zen2019libritts} dataset, and the training criterion is detailed in \cite{miao2022language}.

\noindent
\textit{2) Speech synthesis.} The frame-wise content features, F0, and utterance-level source speaker vector are passed to the HiFi-GAN neural vocoder \cite{kong2020hifi} after the operation of upsampling and concatenation to synthesize speech.
The HiFi-GAN model was trained on \textit{LibriTTS-train-clean-100} \cite{zen2019libritts}, denoted as  monolingual HiFi-GAN.

An extra step called \textit{Speaker vector anonymization} is included in the anonymization stage \cite{tomashenko2020introducing,tomashenko2021voiceprivacy}.
Given a source speaker vector, cosine distance is used to find the 200 farthest speaker vectors from an external speaker vector pool (\textit{LibriTTS train-other-500}). From these 200 vectors, 100 vectors are randomly selected and their average is used as the anonymized speaker vector \cite{Srivastava2020DesignCF}. Then, the content features, F0 and anonymized speaker vector instead of the source speaker vector are used to generate the anonymized speech.

\section{Analysis of Performance Bottleneck}
Although the above SAS performed reasonably well on the unseen Mandarin data, the anonymization performance was below the theoretical optimum. Furthermore, the contents of the anonymized speech were distorted by the SAS. It was shown that the CER of the anonymized speech reached 18.92\%, which was much higher than the CER of 10.36\% on the original data. 
Also, when the output speech was produced without actually anonymizing the speaker vector (i.e., resynthesis), its CER was still 14.81\%.
However, resynthesized speech from an ideal SAS should obtain a similar CER to the original speech. 

The above results motivated us to analyze the performance bottleneck of our SAS under unseen circumstances. Here we examine the components in the system one by one and present techniques to alleviate the bottlenecks. 
Although we only consider Mandarin speech as the unseen test data, the analyses are expected to apply to other unseen languages as well. 

\subsection{Robustness of HiFi-GAN}
The CER gap between the original (10.36\%) and resynthesized speech data (14.81\%) suggests that the speech contents in the resynthesized speech may have been distorted. A similar finding has been reported on the language-dependent VPC baseline \cite{champion:hal-02995855}. 
Since the system extracted features and re-synthesized the speech waveform without anonymizing the speaker representations, the increased CER is only caused by the content feature extractor and vocoder. 

We hypothesize that the HiFi-GAN vocoder is one bottleneck\footnote{We also investigated a multilingual-trained SSL-based soft content encoder but no improvements were observed.}. It is known that to build a neural vocoder that generalizes well to unseen speakers in unseen languages, the training dataset has to cover diverse speakers and languages \cite{Lorenzo-Trueba2019}. However, the HiFi-GAN in our SAS was trained using a subset from LibriTTS with many speakers, but only English-speaking ones. 
To verify the hypothesis, we used a multilingual database \cite{pratap20_interspeech} to train the HiFi-GAN and compared its performance with the HiFi-GAN trained on the monolingual LibriTTS data in the experiments. This comparison is illustrated in Figure~\ref{fig:structures}(a). Note that the multilingual database does not contain Mandarin data, and its details are explained in Section \ref{sec:database}.

\subsection{Language and Channel Mismatch on Speaker Vectors}
Comparing the CERs in the resynthesized (14.81\%) and anonymized (18.92\%) cases, the increased CER is likely due to the anonymized speaker vector since it is the only difference between the resynthesized and anonymized data.
Many studies have shown that speaker vectors contain speaker-unrelated information from the source domain, e.g., channel conditions and lexical contents \cite{raj2019probing, williams2019disentangling}. Because the anonymized vector is composed from the pool of English speaker vectors (i.e., those from \textit{LibriTTS train-other-500} \cite{zen2019libritts}), it may carry irrelevant information pertinent to the English database. Therefore, directly using the anonymized vector on Mandarin data may introduce language, channel, or other types of domain mismatch. 

\begin{figure}[!t]
\centering
 \vspace{-0.5cm}
 \centering
  \includegraphics[width=6cm]{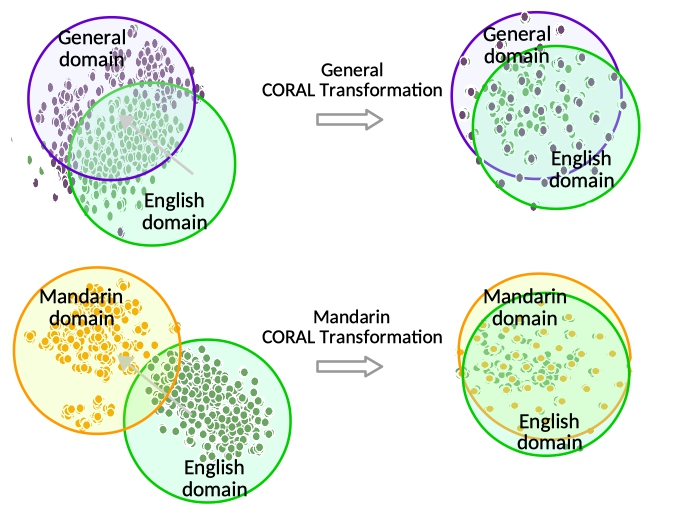}
   \vspace{-0.2cm}
\caption{CORAL transformation on speaker vectors.
}
\label{fig:coral}
 \vspace{-0.6cm}
\end{figure}

To verify this hypothesis,
we use a simple but effective unsupervised domain adaption technique called CORAL \cite{sun2016return}. 
The goal of CORAL is to find a transfer matrix $ \bm{A} $ that can align the feature distributions of the source domain and target domain by minimizing their covariances.
Suppose source-domain English speaker vectors $ \mathcal{D}_S = \lbrace \mathbf n_i \rbrace$,  $\mathbf n_i  \in {\mathbb{R}^{192}} $, 
and the target speaker vectors  $ \mathcal{D}_T = \lbrace \mathbf m_i \rbrace$,  $\mathbf m_i  \in {\mathbb{R}^{192}} $,
here $\mathbf n_i $ and ${ \mathbf m_i}$, are the 192-dimensional vectors extracted from the last projection layer of the ECAPA-TDNN.
After the feature normalization where the speaker vectors $ \mathcal{D}_S$ and $ \mathcal{D}_T$ are normalized to have zero mean and unit standard deviation in each dimension, 
the relationship of source- and target-domain statistics is that mean $ \bm{\mu}_S =  \bm{\mu}_T = 0$ and covariance matrices $  \bm{C}_S \neq  \bm{C}_T$.
The transfer matrix $ \bm{A} $ can be obtained via:
\setlength{\belowdisplayskip}{1pt}
\begin{align*}
\label{eqn:coral}
CORAL( \mathcal{D}_S,  \mathcal{D}_T) = \min\limits_{ \bm{A}} \begin{Vmatrix} {\bm{A}}^T  \bm{C}_S  \bm{A} -  \bm{C}_T \end{Vmatrix} ^2 _F
\end{align*}
where $ \begin{Vmatrix}  . \end{Vmatrix} ^2 _F $ represents the matrix Frobenius norm. 
The detailed procedure to compute the optimal $ \bm{A}^*$ can be found in \cite{sun2016return,cai2010singular}.
Then, the anonymized English speaker vectors transformed by $\bm{A}^*$ are used as the new anonymized speaker vector. 

We consider two cases of CORAL transformations based on the data availability:

\noindent
\textit{\textbf{ General CORAL transformation}}: we assume that Mandarin data are unavailable, and the target-domain speaker vectors are collected from the German, Italian, and Spanish speech data; 

\noindent
\textit{\textbf{ Mandarin CORAL transformation}}: we assume that a few (no more than 100) Mandarin samples that are completely disjoint (unseen-unheard speakers) with the test set are available as the target-domain speaker vectors.

In the first case, where the domain of the test data to be anonymized remains unseen to the SAS, CORAL using the target speaker vectors from multiple languages and channels is expected to alleviate the mismatch between the English vector pool and the vector to be anonymized. 
The second case is an oracle scenario where the mismatch caused by the impure speaker vectors can be reduced to a greater extent when they are transferred to the matched domain.

Figure \ref{fig:coral} plots the t-SNE visualization of speaker vectors before and after the two cases of CORAL transformations.
These speaker vectors are sampled from the different datasets listed at the bottom of Table \ref{tab:datasets}.
It is observed that the original distributions of speaker vectors from the different domains are widely spread due to the language and channel mismatch. The distribution between English and general domains has more overlap than between the English and Mandarin domains. 
Despite the different degrees of overlap, CORAL transformations push the speaker vectors from different domains to move closer.

\section{Experiments}
\label{sec:exc}
\subsection{Evaluation Protocols}
Our experiments followed the evaluation protocols of VPC 2020 \cite{tomashenko2020introducing,tomashenko2021voiceprivacy}. 
The SAS anonymized the Mandarin test trials shared by users to protect the speaker identity while preserving the speech contents. 
To assess how well the speech contents are preserved, CER was computed using a language-matched ASR (\emph{$ASR_\text{eval}$}) as one utility metric. 
The protection of speaker identity was evaluated via one privacy matrix with equal error rates (EERs) of a language-matched ASV evaluation model (\emph{$ASV_\text{eval}$}) in two setups: 

\noindent
\textbf{\textit{Ignorant}}: the attackers have access to the \emph{$ASV_\text{eval}$} and a few unanonymized test trials from the users. With \emph{$ASV_\text{eval}$}, the attackers try to recognize the speaker identity by matching the test trials with the unanonymized data (i.e., enrollment data). However, they are unaware that the test trials have been anonymized;

\noindent
\textbf{\textit{Lazy-informed}}: the attackers have the same resources as in \textit{Ignorant}. Besides these, they know that the test trials have been anonymized using the anonymization algorithm and CORAL transformation, but are unsure of the detailed parameters (e.g., which speaker vectors are selected to compose the anonymized vector). 
The attackers anonymize the enrollment data with their knowledge of SAS and use them to recognize the test trials.

This study included another two setups as references: \textbf{\textit{Unprotected}}: the attackers directly recognize the unanonymized test trials using the original enrollment data and \emph{$ASV_\text{eval}$}; 
\textbf{\textit{Resynthesized}}: similar to \textit{Unprotected}, but the test trials are resynthesized by the SAS using the original speaker vector.
Both setups simulate the case where speaker identity is unprotected, but \textit{Resynthesized} further examines how the feature extractors and vocoder degrade the utility and privacy in the resynthesis process \cite{champion:hal-02995855}.

An ideal SAS should achieve high EERs (close to 50\%) in both \textit{Ignorant} and \textit{Lazy-informed} setups. 
The EERs in \textit{Resynthesized} should be as low as those from \textit{Unprotected}. 
Meanwhile, the CERs for all setups should be low.

\begin{table}[!t]
\vspace{-0.3cm}
\footnotesize
\caption{The datasets used in the different models.}
\label{tab:datasets}
\vspace{-0.2cm}
\begin{tabular}{|ll|l|}
\hline
\multicolumn{2}{|l|}{Model }                                                                    & Dataset                      \\ \hline
\multicolumn{1}{|l|}{\multirow{2}{*}{\rotatebox{90}{Training}} }                 & Monolingual HiFi-GAN                   & LibriTTS train-clean-100    \\ \cline{2-3} 
\multicolumn{1}{|c|}{}                                & \multirow{2}{*}{Multilingual HiFi-GAN} &         German \& Italian \& Spanish         \\ \cline{3-3} 
\multicolumn{1}{|c|}{}                                &                                        &      LibriTTS train-clean-100                    \\ \hline \hline
\multicolumn{1}{|l|}{\multirow{2}{*}{\rotatebox{90}{Tran.}}}     & \multicolumn{1}{l|}{General CORAL}               & \multicolumn{1}{l|}{German \& Italian \& Spanish   } \\ \cline{2-3} 
\multicolumn{1}{|l|}{}                                & \multicolumn{1}{l|}{Mandarin CORAL}               & \multicolumn{1}{l|}{AISHELL-3-test-left}    \\ \hline 
\end{tabular}
\vspace{-0.5cm}
\end{table}

\begin{figure*}[t]
\vspace{-0.8cm}
  \centering
  \includegraphics[width=16cm, height=4cm]{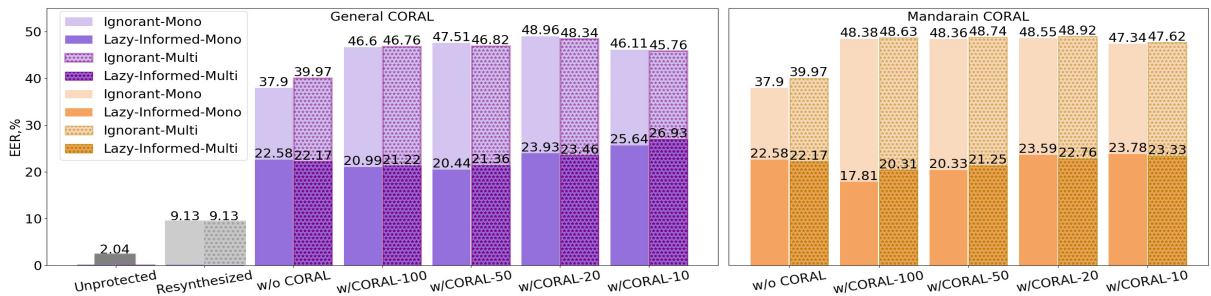}
   \vspace{-0.3cm}
  \caption{EER values for two anonymization systems using monolingual HiFi-GAN or multilingual HiFi-GAN along with different configurations of the general and Mandarin CORAL transformation.}
  \label{fig:eer}
  \vspace{-0.4cm}
\end{figure*}

\subsection{Databases} 
\label{sec:database}
In addition to the standard databases to build the SAS (Section~\ref{sec:ssl-sas}), this study used external data listed in Table \ref{tab:datasets} for the SAS training and CORAL transformation.
The multilingual dataset for HiFi-GAN consists of \textit{LibriTTS-train-clean-100} and subsets that contains  \textit{German}, \textit{Italian}, and \textit{Spanish} data sampled from the Multilingual LibriSpeech corpus \cite{pratap20_interspeech}.
Around 78 hours of clean data were selected for each language. The selection criteria is that the duration of each recording is larger than 10s, and that the signal-to-noise ratio is equal to 100, which is estimated using the WADA SNR algorithm \cite{kim2008robust}.

The Mandarin test trials and enrollment data were sampled from the test set of Mandarin speech corpus \textit{AISHELL-3} \cite{shi21c_interspeech}. 
Specifically, 4,179 trials from 44 speakers were randomly sampled as the test trials, and additional 2 utterances of the same speaker were sampled for enrollment.
They composed 10,120 enrollment-test pairs for the ASV evaluation, which is denoted as \textit{AISHELL-3-test-veri}.
The target-domain vectors for the oracle Mandarin CORAL were extracted from the left data of \textit{AISHELL-3} test set, denoted \textit{AISHELL-3-test-left} in Table \ref{tab:datasets}. Note that there is no speaker or utterance overlap between \textit{AISHELL-3-test-left} and \textit{AISHELL-3-test-veri}. 
The target-domain vectors for the general CORAL were randomly selected from the \textit{German}, \textit{Italian}, and \textit{Spanish} subsets. 
The source-domain vectors for both types of CORAL were randomly selected from the \textit{LibriTTS-train-clean-100}.
Similar to \cite{miao2022language}, the \emph{$ASV_\text{eval}$} model was an ECAPA-TDNN trained on the Mandarin \textit{CN-Celeb-1 \& 2} \cite{li2022cn,fan2020cn} datasets. The \emph{$ASR_\text{eval}$} model was an open-source ASR Transformer  \cite{speechbrain} trained on the Mandarin \textit{AISHELL-1} ASR dataset \cite{aishell_2017}.

\begin{figure}[!t]
  \centering
  \vspace{0.0cm}
  \includegraphics[width=7cm]{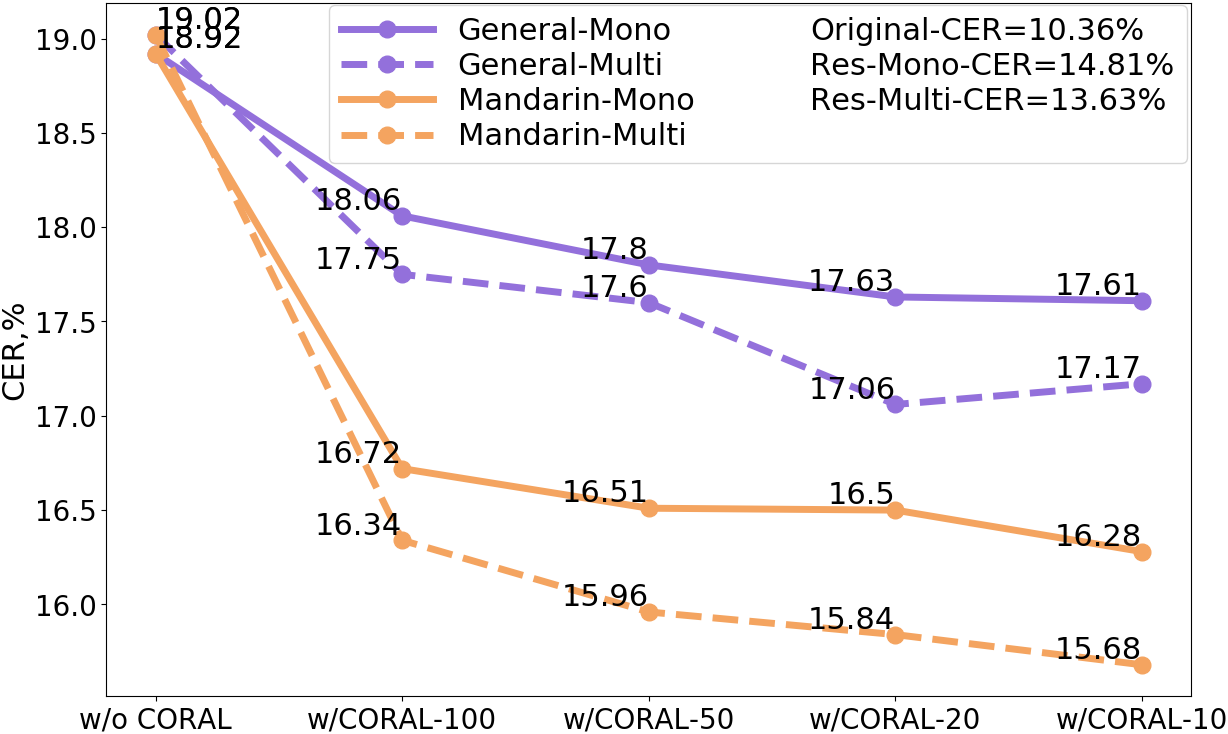}
   \vspace{-0.2cm}
  \caption{CERs of anonymized speech on AISHELL-3-test-veri.}
  \label{fig:cer}
  \vspace{-0.7cm}
\end{figure}

\subsection{Results and analysis}
Our experiments compared the SAS performance on the Mandarin test data by varying the two factors illustrated in Figure~\ref{fig:structures}(b): using monolingual or multilingual data to train the HiFi-GAN vocoder, and whether CORAL was applied on the anonymized speaker vector. 
The EERs from different setups are plotted in Figure~\ref{fig:eer}. 
The CERs are plotted in Figure~\ref{fig:cer}. 
Note that, when CORAL was used, we also analyzed the impact of using different amounts of data to estimate the CORAL transformation matrix. The label ``w/CORAL-$N$'' in the figures indicates using $N$ randomly-chosen speaker vectors from the source and target domain respectively. 
All the results for CORAL transformation are computed over 5 runs of speaker vector selection.

\noindent
\textbf{Monolingual vs. Multilingual HiFi-GAN:} 
In the \textit{Resynthesized} setup, the SAS using multilingual training data for HiFi-GAN achieved a CER of 13.63\%, which is lower than 14.81\% when using monolingual data. Meanwhile, the EERs in both cases are equal to 9.13\%. 
When anonymization was conducted, the SASs using multilingual HiFi-GAN achieved lower CERs than their counterparts using monolingual HiFi-GAN in most of the setups except when no CORAL was used (w/o CORAL, 19.02\% $>$ 18.92\%). 
Specifically, when the Mandarin CORAL was used, the multilingual HiFi-GAN outperformed the monolingual case in terms of CER, no matter how much data was used to estimate the CORAL matrix.

These results indicate that the multilingual data is helpful to obtain a robust HiFi-GAN to preserve the speech contents better. The improvement is expected to be larger when the domain mismatch on the anonymized speaker vectors is reduced using CORAL. The EERs were roughly similar no matter which HiFi-GAN was used. Therefore, the benefit of using the multilingual HiFi-GAN is mainly on the better preservation of the speech contents, rather than better protection of the speaker identity.

\noindent
\textbf{CORAL:}  
Compared to ``w/o CORAL", all the EERs on the \textit{Ignorant} condition of ``w/CORAL-*'' were successfully increased regardless of the choice of HiFi-GAN training data, CORAL types, and amount of data for CORAL matrix estimation. Similarly, the CERs were significantly decreased after applying CORAL. 
These results suggest that the mismatch from the anonymized speaker vectors severely affected the SAS, and CORAL is effective to reduce various types of mismatches (e.g., language and channel) between different datasets/domains. 

For the different configurations for CORAL, we first observed that the oracle Mandarin CORAL performed better on CERs than the general CORAL. However, their differences on EERs are not obvious. This indicates that the SAS performance on speech content preservation is more sensitive to the mismatch of anonymized vectors than the CORAL configurations. For speaker identity protection, using the general CORAL is sufficient. 
Interestingly, unlike DNN-based methods, using larger $N$ to estimate the CORAL matrix did not constantly improve the results. Using 20 samples (w/CORAL-20) generally performed well in terms of EER and CER.
The reason is that users and attackers randomly choose speaker vectors from the target domain individually to approximate CORAL transformation matrices. 
These matrices can be very different if the number of the speaker vectors $N$ is relatively small, which increases the randomness of the new anonymized speaker vectors used by the users and attackers. Therefore, speaker identity information can be protected better. 

Considering that the users may prefer to choose smaller $N$ to protect their privacy, while an attacker may be interested to use larger $N$ to find a more precise CORAL transform matrix on the \textit{Lazy-informed} condition, we then set $N=10$ for users and $N=100$ for attackers to compute the CORAL matrix independently, denoted CORAL-10-100.
The results from Table \ref{tab:diff-N} show that the EERs of CORAL-10-100 are lower than those of CORAL-10 for both general and Mandarin cases in an acceptable level. Furthermore, "General CORAL-100-10" still performs better than "w/o CORAL".  

\begin{table}[!t]
\centering
\vspace{-4mm}
\caption{EER values on \textit{Lazy-informed} condition for the SAS using Multiligual HiFi-GAN and different CORAL configurations.}
\vspace{-2mm}
    \footnotesize
  \label{tab:diff-N}
\begin{tabular}{l|c|c|c}
EER(\%) & w/o CORAL & CORAL-10 & CORAL-10-100 \\ \hline           
General & 22.17 & 26.93 & 25.93 \\ 
Mandarin & 22.17 & 23.33 & 21.66 \\
\end{tabular}
\vspace{-7mm}
\end{table}

\section{Conclusions}
This paper analyzed the previously proposed SSL-based SAS under unseen conditions. 
Two hypotheses, which are that the performance bottleneck exists in the HiFi-GAN and in anonymized speaker vectors, were presented and experimentally verified. The results indicate that increasing the language diversity for the HiFi-GAN benefits the preservation of speech contents. The mismatch on the anonymized speaker vectors severely affect the SAS. The SAS using multilingual HiFi-GAN and CORAL strategy easily outperforms the previous SAS using monolingual HiFi-GAN on both privacy and utility.

\noindent
\textbf{Acknowledgements}
This study is supported by JST CREST Grants (JPMJCR18A6 and JPMJCR20D3), MEXT KAKENHI Grants (21K17775, 21H04906, 21K11951, 18H04112),
and the VoicePersonal project (ANR-18-JSTS-0001).

\bibliographystyle{IEEEtran}

\bibliography{main}
\end{document}